\documentclass[twocolumn,epsfig,aps,prb]{revtex4}

\begin{document}

\title{Plasmon-polaritons on the surface of a pseudosphere.}

\author{Igor I. Smolyaninov, Quirino Balzano, Christopher C. Davis}
\affiliation{Department of Electrical and Computer Engineering, University of Maryland, College Park, MD 20742, USA}

\date{\today}

\begin{abstract}
Plasmon-polariton behavior on the surfaces of constant negative curvature is considered. Analytical solutions for the eigenfrequencies and eigenfunctions are found, which describe resonant behavior of metal nanotips and nanoholes. These resonances may be used to improve performance of metal tips and nanoapertures in various scanning probe microscopy and surface-enhanced spectroscopy techniques.   
\end{abstract}


\maketitle

Surface plasmon-polaritons (SP) \cite{1,2} play an important role in the properties of metallic nanostructures. Optical field enhancement due to SP resonances of nanostructures (such as metal tips or nanoholes) is the key factor, which defines performance of numerous scanning probe microscopy and surface-enhanced spectroscopy techniques. Unfortunately, it is often very difficult to predict or anticipate what kind of metal tip would produce the best image contrast \cite{3}, or what shape of a nanohole in metal film would produce the best optical throughput. Trial and error with many different shapes is often the only way to produce good results in the experiment and/or in the computer simulation. The reason for this situation is that only a few nanostructure geometries have known analytical solutions, which may be used in designing a good scanning probe or sensor surface shapes. At the moment, these analytically-solvable geometries are basically limited to the planar, spherical and cylindrical cases. Unfortunately, these geometries may serve only as very poor approximations of the shapes of many real-life scanning probe tips and nanoholes. It is easy to notice that in all the known analytically-solvable cases mentioned above the surface curvature is either positive or zero. On the other hand, the best results in scanning probe microscopy are often obtained with metal tips, which have negative curvature (Fig.1(a)). The opening of real-life nanoholes has negative curvature too (Fig.1(b)). Thus, understanding of the properties of plasmon-polaritons on the so-called hyperbolic surfaces (surfaces of negative curvature) is essential in understanding the optics of nanostructures. 

In this paper we consider the properties of surface plasmon-polaritons, which propagate over a surface of a pseudosphere \cite{4}. Pseudosphere (or anti-sphere, shown in Fig.2(a)) is the simplest surface exhibiting hyperbolic geometry. Its Gaussian curvature is negative and constant everywhere. We are going to show that the SP resonances of the pseudosphere may play no lesser role in nanophotonics than the well-established role of spherical plasmon resonance of metal nanoparticles \cite{1}. 

The pseudosphere is produced as a surface of revolution of the tractrix (Fig.2(b)), which is the curve in which the distance $a$ from the point of contact of a tangent to the point where it cuts its asymptote is constant. In Cartesian coordinates the equation of tractrix is   

\begin{equation}
\label{eq1}
z=aln(\frac{a}{x}+(\frac{a^2}{x^2}-1)^{1/2})-a(1-\frac{x^2}{a^2})^{1/2}
\end{equation}

The Gaussian curvature of the pseudosphere is $K=-1/a^2$ everywhere on its surface (so it may be understood as a sphere of imaginary radius $ia$). It is quite apparent from comparison of Figs.1 and 2 that the pseudospherical shape may be a good approximation of the shape of a nanotip or a nanohole. On the other hand, this surface possesses very nice analytical properties, which makes it very convenient and easy to analyze. 

The surface of the pseudosphere may be understood as a sector (a horocyclic sector \cite{4} in more precise non-Euclidian geometry terms) of the hyperbolic plane, an infinite surface with the metrics     

\begin{equation}
\label{eq2}
ds^2=R^2(d\Theta^2+sinh^2\Theta d\Phi^2),
\end{equation}

which (as shown by Hilbert) cannot be entirely fitted isometrically into the Euclidean space $R^3$. If the edges of the horocyclic sector are glued together along the geodesics CC\dag (Fig.2(a)), the pseudospherical shape is obtained. Thus, the small-scale local metrics on the surface of the pseudosphere is described by equation (2). The physics of quantum fields in hyperbolic spaces of various dimensions (such as the anti-de Sitter space) is a topic of very active current research \cite{5,6,7}. Since nonlinear optics of surface plasmon-polaritons provides an interesting toy model of quantum gravity in a curved space-time background \cite{8}, the study of plasmon optics on a pseudosphere may have important field-theoretical implications, in addition to practical benefits of building good scanning probes and nanosensors.

Before we proceed with writing down the Maxwell equations for surface plasmon-polaritons on a pseudosphere, let us discuss and emphasize important analytical properties of the two-dimensional optics of SPs on a hyperbolic plane. Even though the surface of the pseudosphere represents only a sector of the plane, this approximation is useful when the SP wavelength is small. In this (ray-optics) approximation the SPs behave as free classical particles on a hyperbolic plane. They move freely without scattering along various geodesics (the straight lines of the non-Euclidian geometry on the hyperbolic plane). You do not need to solve anything in this approximation! The motion of the surface wave on the curved surface of the hyperbolic plane is a free motion, because the curvature is the same everywhere. Since $\phi=const$ curves on the surface of the pseudosphere in Fig.2 are geodesics, a short-wavelength SP wave packet launched towards the apex of the pseudosphere should freely move straight down the apex without scattering, indefinitely. This is a robust topological property of the pseudosphere, which is extremely important for the effect of optical field enhancement at the apex of the tip, or deep inside the nanohole. Such robust and predictable shapes are very desirable in scanning probe microscopy and spectroscopy.         

Unfortunately, this ray-optics approximation breaks down when the tip radius becomes comparable with the SP wavelength. We must solve Maxwell equations in order to find the exact field distribution near the tip apex. Luckily, nice geometrical properties of the pseudosphere allow us to find exact analytical solution of this problem. Let us introduce rectangular curvilinear coordinates $(\alpha, \theta, \phi)$ as shown in fig.2, so that the distance from a point on the pseudosphere to the $z$-axis is $\rho=asin\theta$, and the radius of curvature of the $\phi=const$ tractrix  is $r=a/tan\theta$ (where $\vec{r}$ connects the point on the tractrix with the center of the osculating circle at this point; note that the locations of all the osculating circles form the evolute of the tractrix, which is called the catenary, see fig.2(b)). The third coordinate in the orthogonal set, which is normal to the surface of the pseudosphere is defined in the vicinity of the pseudosphere surface as $ad\alpha$, since the tractrix is orthogonal to a set of circles of constant radius $a$ whose centers are located on the $z$-axis, as shown in Fig.2(b). Thus, the spatial metrics in this system of coordinates takes the form 

\begin{equation}
\label{eq3}
ds^2=a^2d\alpha^2+a^2\frac{cos^2\theta}{sin^2\theta}d\theta^2+a^2sin^2\theta d\phi^2
\end{equation}

We should emphasize that the so introduced global angular coordinates $(\theta, \phi)$ on the surface of the pseudosphere and the local angular coordinates $(\Theta, \Phi)$ in equation (2) are not the same. The local $(\Theta, \Phi)$ coordinates provide good representation of the metrics of the pseudosphere surface only when $R<<\rho(\theta)$.

The Maxwell equations in the curvilinear $(\alpha, \theta, \phi)$ coordinates 

\begin{equation}
\label{eq4}
\vec{\nabla}\times\vec{\nabla}\times\vec{H}=\frac{\omega^2\epsilon(\omega)}{c^2}\vec{H},
\end{equation}

(where $\epsilon(\omega)$ is the dielectric constant of metal), and the continuity of $D_{\alpha}$ on the metal surface can be used in order to find the dispersion law of surface plasmon-polaritons on the surface of the pseudosphere. Introducing a new variable $\tau$ as $\tau=-ln(sin\theta)$, which has a clear physical meaning of the normalized arc length of the tractrix $\phi=const$, the Maxwell equations (4) lead to  

\begin{equation}
\label{eq5}
\frac{1}{a^2}\frac{\partial^2H_{\phi}}{\partial\alpha^2}+\frac{\omega^2\epsilon(\omega)}{c^2}H_{\phi}+\frac{1}{a^2}(\frac{\partial^2H_{\phi}}{\partial\tau^2}-\frac{\partial H_{\phi}}{\partial\tau}+e^{2\tau}\frac{\partial^2H_{\phi }}{\partial\phi^2})=0
\end{equation}

We are going to search for surface-wave-like solutions with a well-defined angular momentum $n$ of the form

\begin{equation}
\label{eq6}
H_{\phi}=H_0e^{-\kappa\alpha}e^{in\phi}e^{ik\tau}
\end{equation}

In the most interesting case of $n=0$ modes (the modes that correspond to the rays moving straight down the apex), equation (6) gives

\begin{equation}
\label{eq7}
k^2-ik-(\kappa_0^2+\frac{a^2\omega^2}{c^2})=0
\end{equation}

in vacuum, and 

\begin{equation}
\label{eq8}
k^2-ik-(\kappa_m^2+\frac{a^2\omega^2\epsilon(\omega)}{c^2})=0
\end{equation}

in metal, where $\kappa_0$ and $\kappa_m$ are the exponents of the field decay perpendicular to the interface in vacuum and metal, respectively. The boundary condition $\kappa_0=-\kappa_m\epsilon(\omega)$ leads to the following solution for $k$ and $\kappa_0$:

\begin{equation}
\label{eq9}
k=\pm(\frac{a^2\omega^2\epsilon}{c^2(\epsilon+1)}-\frac{1}{4})^{1/2}+\frac{i}{2},
\end{equation}

\begin{equation}
\label{eq10}
\kappa_0=-\frac{a^2\omega^2}{c^2(\epsilon+1)}
\end{equation}

In the Drude model, in which $\epsilon(\omega)=1-\omega^2_p/\omega^2$ is real, the surface-wave-like solution with $n=0$ exists if $\epsilon<-1$. In vacuum the SP field is 

\begin{equation}
\label{eq11}
H_{\phi}=H_0e^{-\kappa_0\alpha}e^{-\tau/2}e^{ik\ast\tau},
\end{equation}

where we have introduced $k^{\ast}=k-i/2$. The SP dispersion law of this mode looks like

\begin{equation}
\label{eq12}
k^{\ast}=\pm(\frac{a^2\omega^2(1-\omega_p^2/\omega^2)}{c^2(2-\omega_p^2/\omega^2)}-\frac{1}{4})^{1/2}
\end{equation}

It is easy to see that in the $\omega<<\omega_p$ limit, the gap appears in the SP spectrum, as usually happens on the surfaces of negative curvature \cite{5}. In the low-frequency limit  

\begin{equation}
\label{eq13}
k^{\ast}=\pm(\frac{a^2\omega^2}{c^2}-\frac{1}{4})^{1/2},
\end{equation}

where the frequency gap width is related to the curvature as $\Delta\omega=c/2a$.

In the purely theoretical case of a complete infinite pseudosphere surface the spectrum of surface plasmon-polaritons is continuous. However, any practical STM tip is not infinitely sharp (Fig.1(a)). Real tips may be approximated by a portion of the pseudosphere surface, which is cut at some $\rho_{min}$, and hence some $\tau_{max}$. In such case the tip would exhibit a set of resonant eigenfrequencies, which correspond to the excitation of standing SP waves on the tip surface. The resonant condition is given by $k^{\ast}\tau_{max}=\pi m$, where $m$ is integer. While quite simple and straightforward in terms of the normalized pseudosphere arc length $\tau_{max}$ and $k^{\ast}$, this condition is not as evident in terms of the tip length $z_{max}$ and light frequency $\omega$ (see eqs.(1) and (12)), which explains why the reasons for very good performance of some scanning probe tips (compared to very bad one of others) were unclear in local scanning probe spectroscopy measurements \cite{3}. Until now the only good experimental strategy was to try large number of tips at random, until the good one is found. We believe that this situation may be radically changed now. Instead of somewhat cumbersome equation (1), a simpler looking parametrization may be used via introduction of a new variable $\gamma$ as $cosh\gamma=a/\rho$, so that $\tau=ln(cosh\gamma)$, and 

\begin{equation}
\label{eq14}
z=a(\gamma-tanh\gamma)
\end{equation}

Thus, analytical calculation of tip resonances becomes a very straightforward task. An example of a field intensity distribution in one of these resonant modes with $m=4$ is shown in Fig.3. According to eq.(11), the field intensity on the surface is proportional to $(\rho/a)sin^2(k^{\ast}ln(a/\rho))$, so the SP intensity in SP field maxima attenuates roughly as $\rho/a$. In effect, a pseudospherical tip with a pseudo-radius $a>\lambda/4\pi$ (thus, of the order of a few hundreds of nanometers) behaves as a good optical nanoantenna in the $\sim\lambda$ wavelength range. If one of the eigenfrequencies of such a nanoantenna is matched with the frequency of the spherical plasmon resonance \cite{1,2} at $\omega_p/3^{1/2}$, and a spherical metal nanoparticle is placed (or formed) at the apex of the pseudosphere, a very efficient way of delivering optical energy to a nanoscale region of space will result. Such plasmon nanolenses are being actively developed at the moment \cite{9,10}. Note that the limitation  $a>\lambda/4\pi$ on the pseudo-radius of the resonant pseudospherical nanoantenna originates from the frequency gap $\Delta\omega=c/2a$ in the SP spectrum due to negative curvature.

In conclusion, plasmon-polariton behavior on the surfaces of constant negative curvature has been considered. Analytical solutions for the eigenfrequencies and eigenfunctions have been obtained, which describe resonant behavior of metal nanotips and nanoholes. These resonances may be used to improve performance of metal tips and nanoapertures in various scanning probe microscopy and surface-enhanced spectroscopy techniques.

This work has been supported in part by the NSF grant ECS-0304046.

\begin{figure}
\begin{center}
\end{center}
\caption{(a) Electron microscope photo of an etched metal tip of a scanning tunneling microscope. (b) 3D view of a nanohole array produced in a 100 nm thick gold film using focused ion beam milling. The nanohole periodicity is 500 nm. Both surfaces exhibit negative curvature around the tip and nanoholes, respectively.}
\end{figure}

\begin{figure}
\begin{center}
\end{center}
\caption{(a) Pseudosphere, which is obtained as a surface of revolution of a tractrix (b), may be used as a good approximation of the shape of a metal tip of a scanning probe microscope, or the shape of a nanohole in a metal film. Also shown in (b) is a catenary, which is an evolute of the tractrix.}
\end{figure}

\begin{figure}
\begin{center}
\end{center}
\caption{SP field intensity distribution of the $m=4$ resonant mode on the surface of a pseudosphere. The pseudosphere boundary is shown by dots.}
\end{figure}

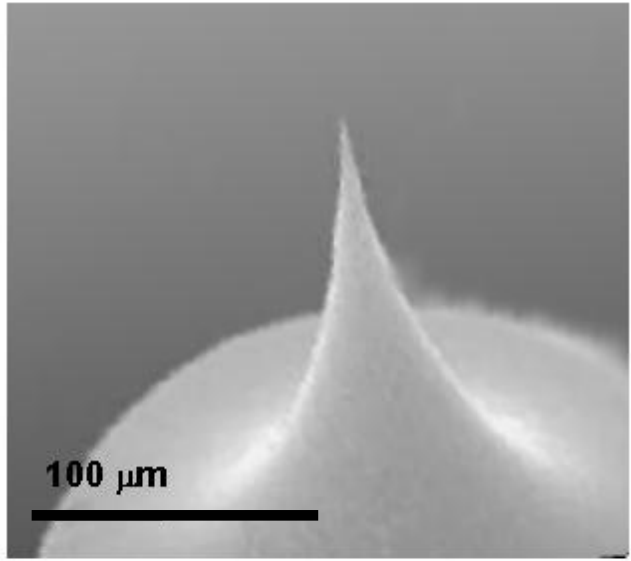
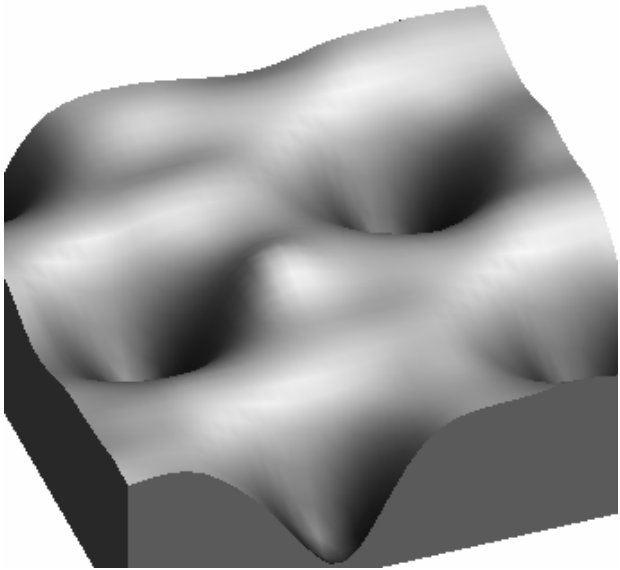
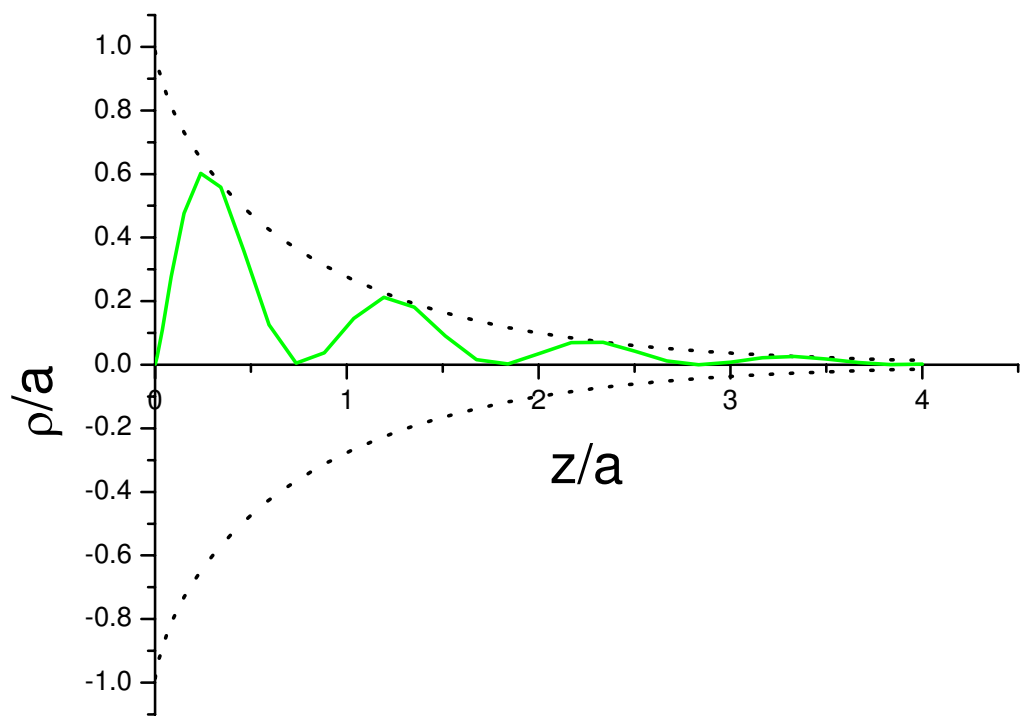
\end{document}